%% file: thz-arxiv.tex
\documentclass[10pt]{IEEEtran}
\usepackage{url}
\usepackage[utf8]{inputenc}
\usepackage{xcolor}
\usepackage{amsmath}
\usepackage{amssymb}

\usepackage[acronyms,nonumberlist,nopostdot,nomain,nogroupskip]{glossaries}
\usepackage{tablefootnote}
\usepackage{booktabs}

\usepackage{tabularx}

\usepackage{tikz}
\usepackage{pgfplots}
\pgfplotsset{compat=newest}
\pgfplotsset{plot coordinates/math parser=false}
\newlength\fheight
\newlength\fwidth
\usetikzlibrary{plotmarks,patterns,decorations.pathreplacing,backgrounds,calc,arrows,arrows.meta,spy,matrix}
\usepgfplotslibrary{patchplots,groupplots}
\usepackage{tikzscale}
\usepackage{hyperref}

\newif\ifexttikz
\exttikzfalse

\ifexttikz
	\usetikzlibrary{external}
\fi

\usepackage{multirow}

\usepackage[font=footnotesize]{subcaption}
\usepackage[font=footnotesize]{caption}

\usepackage{mathtools}

\newacronym{6g}{6G}{sixth generation}
\newacronym{3gpp}{3GPP}{3rd Generation Partnership Project}
\newacronym{adc}{ADC}{Analog to Digital Converter}
\newacronym{5g}{5G}{5th generation}
\newacronym{aimd}{AIMD}{Additive Increase Multiplicative Decrease}
\newacronym{am}{AM}{Acknowledged Mode}
\newacronym{amc}{AMC}{Adaptive Modulation and Coding}
\newacronym{aqm}{AQM}{Active Queue Management}
\newacronym{awgn}{AGWN}{Additive White Gaussian Noise}
\newacronym{balia}{BALIA}{Balanced Link Adaptation}
\newacronym{bdp}{BDP}{Bandwidth-Delay Product}
\newacronym{bf}{BF}{Beamforming}
\newacronym{cc}{CC}{Congestion Control}
\newacronym{cdf}{CDF}{Cumulative Distribution Function}
\newacronym{cn}{CN}{Core Network}
\newacronym{cqi}{CQI}{Channel Quality Information}
\newacronym{cp}{CP}{Control Plane}
\newacronym{csirs}{CSI-RS}{Channel State Information - Reference Signal}
\newacronym{dc}{DC}{Dual Connectivity}
\newacronym{dce}{DCE}{Direct Code Execution}
\newacronym{dci}{DCI}{Downlink Control Information}
\newacronym{dl}{DL}{Downlink}
\newacronym{dmr}{DMR}{Deadline Miss Ratio}
\newacronym{dmrs}{DMRS}{DeModulation Reference Signal}
\newacronym{e2e}{E2E}{End-to-End}
\newacronym{ecn}{ECN}{Explicit Congestion Notification}
\newacronym{edf}{EDF}{Earliest Deadline First}
\newacronym{enb}{eNB}{evolved Node Base}
\newacronym{epc}{EPC}{Evolved Packet Core}
\newacronym{es}{ES}{Edge Server}
\newacronym{fdma}{FDMA}{Frequency Division Multiple Access}
\newacronym{fdd}{FDD}{Frequency Division Duplexing}
\newacronym[firstplural=Radio Access Technologies (RATs)]{rat}{RAT}{Radio Access Technology}
\newacronym{fs}{FS}{Fast Switching}
\newacronym{ftp}{FTP}{File Transfer Protocol}
\newacronym{gnb}{gNB}{Next Generation Node Base}
\newacronym{harq}{HARQ}{Hybrid Automatic Repeat reQuest}
\newacronym{hetnet}{HetNet}{Heterogeneous Network}
\newacronym{hh}{HH}{Hard Handover}
\newacronym{hol}{HOL}{Head-of-Line}
\newacronym{ia}{IA}{Initial Access}
\newacronym{imt}{IMT}{International Mobile Telecommunication}
\newacronym{iot}{IoT}{Internet of Things}
\newacronym{los}{LOS}{Line-of-Sight}
\newacronym{lte}{LTE}{Long Term Evolution}
\newacronym{m2m}{M2M}{Machine to Machine}
\newacronym{mac}{MAC}{Medium Access Control}
\newacronym{mc}{MC}{Multi-Connectivity}
\newacronym{mcs}{MCS}{Modulation and Coding Scheme}
\newacronym{mec}{MEC}{Mobile Edge Cloud}
\newacronym{mi}{MI}{Mutual Information}
\newacronym{mimo}{MIMO}{Multiple Input, Multiple Output}
\newacronym{mmwave}{mmWave}{millimeter wave}
\newacronym{mptcp}{MPTCP}{Multipath TCP}
\newacronym{mr}{MR}{Maximum Rate}
\newacronym{mss}{MSS}{Maximum Segment Size}
\newacronym{mtd}{MTD}{Machine-Type Device}
\newacronym{mtu}{MTU}{Maximum Transmission Unit}
\newacronym{nfv}{NFV}{Network Function Virtualization}
\newacronym{nlos}{NLOS}{Non-Line-of-Sight}
\newacronym{nr}{NR}{New Radio}
\newacronym{ofdm}{OFDM}{Orthogonal Frequency Division Multiplexing}
\newacronym{pdcch}{PDCCH}{Physical Downlink Control Channel}
\newacronym{pdcp}{PDCP}{Packet Data Convergence Protocol}
\newacronym{pdsch}{PDSCH}{Physical Downlink Shared Channel}
\newacronym{pdu}{PDU}{Packet Data Unit}
\newacronym{pf}{PF}{Proportional Fair}
\newacronym{pgw}{PGW}{Packet Gateway}
\newacronym{phy}{PHY}{Physical}
\newacronym{pbch}{PBCH}{Physical Broadcast Channel}
\newacronym[plural=\gls{mme}s,firstplural=Mobility Management Entities (MMEs)]{mme}{MME}{Mobility Management Entity}
\newacronym{prb}{PRB}{Physical Resource Block}
\newacronym{pss}{PSS}{Primary Synchronization Signal}
\newacronym{pucch}{PUCCH}{Physical Uplink Control Channel}
\newacronym{pusch}{PUSCH}{Physical Uplink Shared Channel}
\newacronym{rach}{RACH}{Random Access Channel}
\newacronym{ran}{RAN}{Radio Access Network}
\newacronym{red}{RED}{Random Early Detection}
\newacronym{rf}{RF}{Radio Frequency}
\newacronym{rlc}{RLC}{Radio Link Control}
\newacronym{rlf}{RLF}{Radio Link Failure}
\newacronym{rrc}{RRC}{Radio Resource Control}
\newacronym{rrm}{RRM}{Radio Resource Management}
\newacronym{rr}{RR}{Round Robin}
\newacronym{rs}{RS}{Remote Server}
\newacronym{rsrp}{RSRP}{Reference Signal Received Power}
\newacronym{rss}{RSS}{Received Signal Strength}
\newacronym{rtt}{RTT}{Round Trip Time}
\newacronym{rw}{RW}{Receive Window}
\newacronym{rx}{RX}{Receiver}
\newacronym{sa}{SA}{standalone}
\newacronym{sack}{SACK}{Selective Acknowledgment}
\newacronym{sap}{SAP}{Service Access Point}
\newacronym{sch}{SCH}{Secondary Cell Handover}
\newacronym{scoot}{SCOOT}{Split Cycle Offset Optimization Technique}
\newacronym{sdma}{SDMA}{Spatial Division Multiple Access}
\newacronym{sinr}{SINR}{Signal to Interference plus Noise Ratio}
\newacronym{sm}{SM}{Saturation Mode}
\newacronym{snr}{SNR}{Signal-to-Noise-Ratio}
\newacronym{son}{SON}{Self-Organizing Network}
\newacronym{ss}{SS}{Synchronization Signal}
\newacronym{srs}{SRS}{Sounding Reference Signal}
\newacronym{sss}{SSS}{Secondary Synchronization Signal}
\newacronym{tb}{TB}{Transport Block}
\newacronym{tcp}{TCP}{Transmission Control Protocol}
\newacronym{tdd}{TDD}{Time Division Duplexing}
\newacronym{tdma}{TDMA}{Time Division Multiple Access}
\newacronym{tfl}{TfL}{Transport for London}
\newacronym{tm}{TM}{Transparent Mode}
\newacronym{trp}{TRP}{Transmitter Receiver Pair}
\newacronym{tti}{TTI}{Transmission Time Interval}
\newacronym{ttt}{TTT}{Time-to-Trigger}
\newacronym{tx}{TX}{Transmitter}
\newacronym{ue}{UE}{User Equipment}
\newacronym{ul}{UL}{Uplink}
\newacronym{uml}{UML}{Unified Modeling Language}
\newacronym{um}{UM}{Unacknowledged Mode}
\newacronym{utc}{UTC}{Urban Traffic Control}
\newacronym{vm}{VM}{Virtual Machine}
\newacronym{rsrq}{RSRQ}{Reference Signal Received Quality}
\newacronym{rssi}{RSSI}{Received Signal Strength Indicator}
\newacronym{crs}{CRS}{Cell Reference Signal}
\newacronym{nsa}{NSA}{Non Stand Alone}
\newacronym{mrdc}{MR-DC}{Multi \gls{rat} \gls{dc}}
\newacronym{endc}{EN-DC}{E-UTRAN-\gls{nr} \gls{dc}}
\newacronym{5gc}{5GC}{5G Core}
\newacronym{si}{SI}{Study Item}
\newacronym{iab}{IAB}{Integrated Access and Backhaul}
\newacronym{wf}{WF}{Wired-first}
\newacronym{hqf}{HQF}{Highest-quality-first}
\newacronym{pa}{PA}{Position-aware}
\newacronym{mlr}{MLR}{Maximum-local-rate}
\newacronym{wbf}{WBF}{Wired Bias Function}
\newacronym{mib}{MIB}{Master Information Block}
\newacronym{sib}{SIB}{Secondary Information Block}
\newacronym{kpi}{KPI}{Key Performance Indicator}
\newacronym{ppp}{PPP}{Poisson Point Process}
\newacronym{gtp}{GTP}{GPRS Tunneling Protocol}
\newacronym{amf}{AMF}{Access and Mobility Management Function}
\newacronym{dash}{DASH}{Dynamic Adaptive Streaming over HTTP}
\newacronym{http}{HTTP}{HyperText Transfer Protocol}
\newacronym{qos}{QoS}{Quality of Service}
\newacronym{udp}{UDP}{User Datagram Protocol}
\newacronym{cu}{CU}{Central Unit}
\newacronym{du}{DU}{Distributed Unit}
\newacronym{mt}{MT}{Mobile Termination}
\newacronym{sdap}{SDAP}{Service Data Adaptation Protocol}
\newacronym{tdm}{TDM}{Time Division Multiplexing}
\newacronym{fdm}{FDM}{Frequency Division Multiplexing}
\newacronym{sdm}{SDM}{Space Division Multiplexing}
\newacronym{dag}{DAG}{Directed Acyclic Graph}
\newacronym{st}{ST}{Spanning Tree}
\newacronym{ummimo}{UM-MIMO}{Ultra-massive Multiple Input, Multiple Output}
\newacronym{wlan}{WLAN}{Wireless LAN}
\newacronym{rlnc}{RLNC}{Random Linear Network Coding}
\newacronym{drx}{DRX}{Discontinuous Reception}
\newacronym{cpu}{CPU}{Central Processing Unit}
\newacronym{tls}{TLS}{Transport Layer Security}

\tikzstyle{startstop} = [rectangle, rounded corners, minimum width=2cm, minimum height=0.5cm,text centered, draw=black]
\tikzstyle{io} = [trapezium, trapezium left angle=70, trapezium right angle=110, minimum width=3cm, minimum height=1cm, text centered, draw=black]
\tikzstyle{process} = [rectangle, minimum width=2cm, minimum height=0.5cm, text centered, draw=black, alignb=center]
\tikzstyle{decision} = [ellipse, minimum width=2cm, minimum height=1cm, text centered, draw=black]
\tikzstyle{arrow} = [thick,<->,>=stealth]
\tikzstyle{line} = [thick,>=stealth]
\tikzstyle{darrow} = [thick,<->,>=stealth,dashed]
\tikzstyle{sarrow} = [thick,->,>=stealth]
\tikzstyle{larrow} = [line width=0.1mm,dashdotted,->,>=stealth]

\makeatletter
\def\grd@save@target#1{%
  \def\grd@target{#1}}
\def\grd@save@start#1{%
  \def\grd@start{#1}}
\tikzset{
  grid with coordinates/.style={
    to path={%
      \pgfextra{%
        \edef\grd@@target{(\tikztotarget)}%
        \tikz@scan@one@point\grd@save@target\grd@@target\relax
        \edef\grd@@start{(\tikztostart)}%
        \tikz@scan@one@point\grd@save@start\grd@@start\relax
        \draw[minor help lines] (\tikztostart) grid (\tikztotarget);
        \draw[major help lines] (\tikztostart) grid (\tikztotarget);
        \grd@start
        \pgfmathsetmacro{\grd@xa}{\the\pgf@x/1cm}
        \pgfmathsetmacro{\grd@ya}{\the\pgf@y/1cm}
        \grd@target
        \pgfmathsetmacro{\grd@xb}{\the\pgf@x/1cm}
        \pgfmathsetmacro{\grd@yb}{\the\pgf@y/1cm}
        \pgfmathsetmacro{\grd@xc}{\grd@xa + \pgfkeysvalueof{/tikz/grid with coordinates/major step x}}
        \pgfmathsetmacro{\grd@yc}{\grd@ya + \pgfkeysvalueof{/tikz/grid with coordinates/major step y}}
        \foreach \x in {\grd@xa,\grd@xc,...,\grd@xb}
        \node[anchor=north] at (\x,\grd@ya) {\pgfmathprintnumber{\x}};
        \foreach \y in {\grd@ya,\grd@yc,...,\grd@yb}
        \node[anchor=east] at (\grd@xa,\y) {\pgfmathprintnumber{\y}};
      }
    }
  },
  minor help lines/.style={
    help lines,
    gray,
    line cap =round,
    xstep=\pgfkeysvalueof{/tikz/grid with coordinates/minor step x},
    ystep=\pgfkeysvalueof{/tikz/grid with coordinates/minor step y}
  },
  major help lines/.style={
    help lines,
    line cap =round,
    line width=\pgfkeysvalueof{/tikz/grid with coordinates/major line width},
    xstep=\pgfkeysvalueof{/tikz/grid with coordinates/major step x},
    ystep=\pgfkeysvalueof{/tikz/grid with coordinates/major step y}
  },
  grid with coordinates/.cd,
  minor step x/.initial=.5,
  minor step y/.initial=.2,
  major step x/.initial=1,
  major step y/.initial=1,
  major line width/.initial=1pt,
}
\makeatother

\definecolor{desireRed}{RGB}{230,57,60}%
\definecolor{darkPurple}{RGB}{59,31,43}%
\definecolor{springGreen}{RGB}{37,223,145}%
\definecolor{queenBlue}{RGB}{69,123,157}%
\definecolor{spaceCadet}{RGB}{29,53,87}%

\makeglossaries

\begin{document}

\title{Toward End-to-End, Full-Stack\\6G Terahertz Networks}

\author{\IEEEauthorblockN{Michele Polese, \IEEEmembership{Member, IEEE}, Josep Jornet, \IEEEmembership{Member, IEEE},\\Tommaso Melodia, \IEEEmembership{Fellow, IEEE}, Michele Zorzi, \IEEEmembership{Fellow, IEEE}}
\thanks{
Michele Polese, Josep Jornet and Tommaso Melodia are with the Institute for the Wireless Internet of Things, Northeastern University, Boston, MA, USA. Email: \{m.polese, jmjornet, melodia\}@northeastern.edu.

Michele Zorzi is with the Department of Information Engineering (DEI), University of Padova, Italy. Email: zorzi@dei.unipd.it.
}
}


\flushbottom
\setlength{\parskip}{0ex plus0.1ex}

\maketitle
\glsunset{nr}
\glsunset{lte}

\begin{abstract}
Recent evolutions in semiconductors have brought the terahertz band in the spotlight as an enabler for terabit-per-second communications in 6G networks. Most of the research so far, however, has focused on understanding the physics of terahertz devices, circuitry and propagation, and on studying physical layer solutions. However, integrating this technology in complex mobile networks requires a proper design of the full communication stack, to address link- and system-level challenges related to network setup, management, coordination, energy efficiency, and end-to-end connectivity. This paper provides an overview of the issues that need to be overcome to introduce the terahertz spectrum in mobile networks, from a MAC, network and transport layer perspective, with considerations on the performance of end-to-end data flows on terahertz connections.
\end{abstract}


\begin{picture}(0,0)(10,-320)
\put(0,0){
\put(0,20){\footnotesize This paper has been published on IEEE Communications Magazine. Please cite it as M. Polese, J. M. Jornet, T. Melodia and M. Zorzi, "Toward End-to-End,}
\put(0,10){\footnotesize Full-Stack 6G Terahertz Networks," in IEEE Communications Magazine, vol. 58, no. 11, pp. 48-54, November 2020, doi: 10.1109/MCOM.001.2000224.}
\put(0,0){\tiny \copyright 2020 IEEE. Personal use of this material is permitted. Permission from IEEE must be obtained for all other uses, in any current or future media including reprinting/republishing}
\put(0,-5){\tiny this material for advertising or promotional purposes, creating new collective works, for resale or redistribution to servers or lists, or reuse of any copyrighted component of this work in other works.}
\put(0,-20){\scriptsize }}
\end{picture}

\glsresetall

\begin{figure*}
	\centering
	\includegraphics[width=.95\textwidth]{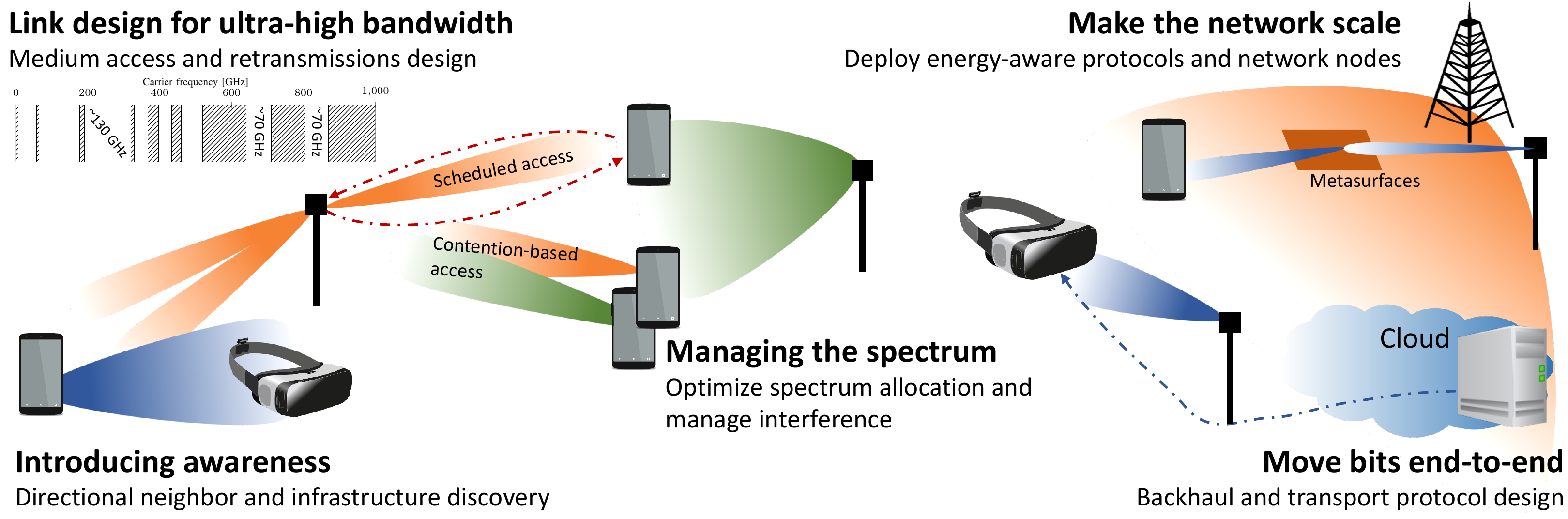}
	\caption{Main design challenges for end-to-end, full-stack terahertz networks.}
	\label{fig:challenges}
\end{figure*}

\section{Introduction}
\label{sec:intro}
Wireless communications will fundamentally shape the innovations of the digital society towards 2030. The unprecedented growth of the mobile traffic has called for the integration of new portions of the spectrum (i.e., between 6 and 52.6 GHz, in the millimeter wave band) in \gls{5g} mobile networks. The 3GPP is already considering an extension to 71 GHz for 3GPP NR, as higher carrier frequencies come with larger bandwidth. For this reason, the terahertz bands are considered as a possible enabler of ultra-high data rates in \gls{6g} networks~\cite{giordani20196g,rappaport2019wireless}. 
The spectrum from 100 GHz to 10 THz, indeed, features wide chunks of untapped bandwidth for communications and sensing. Notably, the IEEE has developed a physical layer that spans 50 GHz of bandwidth, between 275 and 325 GHz.

Terahertz frequencies, however, bring to the extreme the communications and networking challenges of the lower mmWave band. In particular, the harsh propagation environment features a high pathloss, inversely proportional to the square of the wavelength, and thus to the size of a single antenna element, and, in addition, a high molecular absorption in certain frequency bands. Moreover, terahertz signals do not penetrate common materials, and are thus subject to blockage.
Finally, the manufacturing of terahertz devices has been a challenge for years, and only very recent advances in electronics and photonics have enabled portable terahertz equipment~\cite{AKYILDIZ201646}.

Nonetheless, the promise of multi-gigabits-per-second capacity has sparked research efforts to overcome these challenges with novel, efficient, and high-performance physical layer techniques. Several studies have focused on increasing the communication range in macro scenarios~\cite{akyildiz2018combating}, and on signal generation and modulation. Directional antennas are used to mitigate the increased pathloss, as they can focus the power in narrow beams, which increase the link budget, and to enhance the security of wireless links, making eavesdropping more challenging. Furthermore, the small wavelength at terahertz allows many antenna elements to be packed in a small form factor (1024 in 1 mm$^2$ at 1 THz), thus enabling \gls{ummimo} techniques~\cite{AKYILDIZ201646} and array-of-subarrays solutions~\cite{yan2020array}.
Finally, reconfigurable electronic surfaces can act as smart reflectors and overcome blockage in \gls{nlos}~\cite{akyildiz2018combating}. 

These studies, however, mostly focused on modeling, designing and optimizing the physical layer and the RF circuitry. Eventually, only a fraction of the hops between a client and a server will be on terahertz links. Therefore, it is necessary to study and understand the performance of this technology considering the integration in complex, end-to-end networks, where multiple nodes and layers of the protocol stack interact to deliver packets between two applications at the two endpoints of a connection. Moreover, the harsh propagation characteristics of the terahertz band, the limited coverage of a terahertz access point, the directionality, and the huge availability of bandwidth introduce new challenges and potential for the \gls{mac}, network, and transport layers, and may call for a radical re-design of traditional paradigms for user and control planes of wireless networks.

In this paper, we discuss five key areas for the development of end-to-end terahertz networks, summarized in Fig.~\ref{fig:challenges},
by reviewing literature contributions, and providing novel results based on full-stack, end-to-end ns-3 simulations~\cite{hossain2018terasim}.\footnote{The simulation scripts can be found at \url{github.com/mychele/toward-e2e-6g-terahertz-networks}.}
To the best of our knowledge, this is the first contribution that provides a holistic perspective on networking challenges at terahertz frequencies, as prior work mostly focused on the lower portion of the spectrum~\cite{giordani2018tutorial} or on terahertz devices and physical layer~\cite{jornet2011thz,akyildiz2018combating}. Notably, we believe it will be crucial to study mechanisms to:
\begin{itemize}
	\item \textbf{introduce awareness} of neighbors, fixed infrastructure, and channel usage, overcoming \textit{the deafness} introduced by directional communications;
	\item account for the \textbf{ultra-high bandwidth in the link design}, by analyzing the complexity-efficiency trade off of medium access schemes;
	\item \textbf{make the network scale}, with a joint, energy-aware design of the network deployment, considering active and passive nodes;
	\item \textbf{manage the spectrum}, understanding the impact of interference and how the available bandwidth benefits wireless backhaul and multi connectivity;
	\item \textbf{move bits end to end}, analyzing which kinds of transport protocols may provide the best performance.
\end{itemize}
We selected these topics as they represent elements that are impacted by the characteristics of the terahertz channel and deployments, and that can be actually optimized to provide a seamless end-to-end 6G experience on terahertz links. They are discussed in Sections~\ref{sec:awareness}-\ref{sec:e2e}, where, for each area, we first highlight the challenges introduced by terahertz scenarios, and then outline promising research directions. Section~\ref{sec:concl} concludes the paper.

\begin{figure*}
\begin{subfigure}[t]{0.6\textwidth}
	\centering
	\setlength\fwidth{0.7\textwidth}
	\setlength\fheight{0.45\textwidth}
	\input{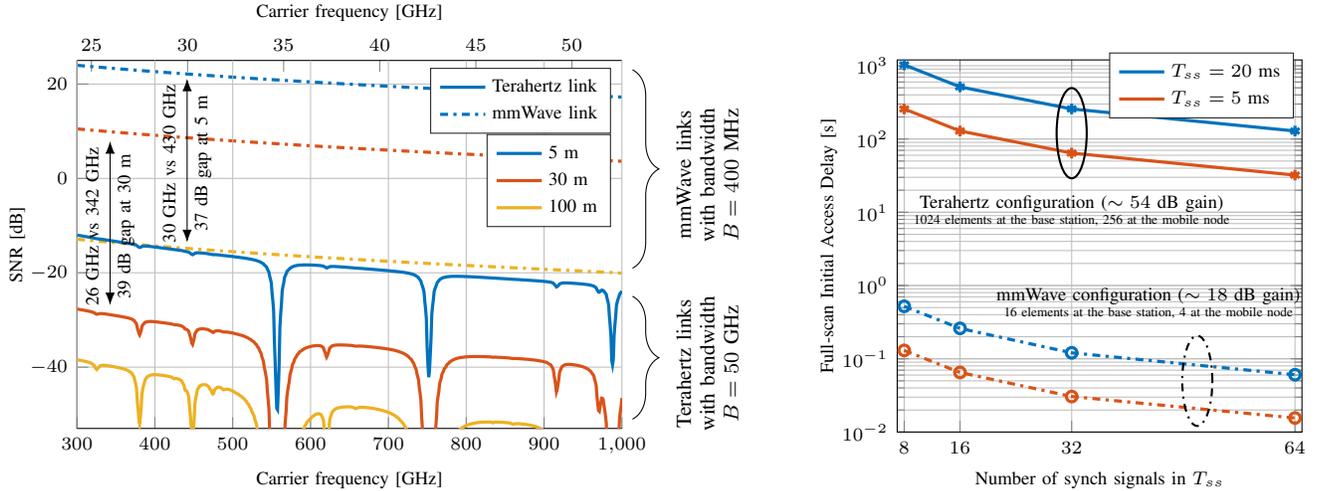}
	\caption{Link budget for \glspl{mmwave} and terahertz links, without beamforming gain. $P_{tx} = 0.5$ W, noise figure $F = 10$ dB.}
	\label{fig:snr}
\end{subfigure}\hfill
\begin{subfigure}[t]{0.39\textwidth}
	\centering
	\setlength\fwidth{0.8\textwidth}
	\setlength\fheight{0.7\textwidth}
	\input{figures/initial_access_nr_thz.tex}
	\caption{Link establishment delay computed with the 3GPP NR parameters and the model from~\cite{giordani2018tutorial}, using antenna arrays that yield a comparable \gls{snr} at the receiver.}
	\label{fig:ia}
\end{subfigure}
	\caption{Link budget and impact of directionality at the \gls{mac} layer.}
	\label{fig:link}
\end{figure*}

\section{Introducing Awareness}
\label{sec:awareness}

Mobile nodes need to gather awareness of the surrounding environment. Even in a random access context, without any coordination, devices still need to be aware of incoming transmissions. Moreover, in a cellular or \gls{wlan} scenario, where mobile users exchange data with a fixed infrastructure (i.e., base stations and access points), each endpoint of the wireless link should have knowledge of the other: in cellular networks, users perform initial access to a base station, which is a gateway to the overall network, and schedules spectrum resources for the connected devices. Finally, they may need to sense the channel to understand if it is busy, and decide to transmit only if idle. Therefore, it is necessary for mobile transceivers to gain awareness of the fixed infrastructure, of their neighbors, and, if needed, of the channel occupancy. 

Traditionally, cellular and \gls{wlan} networks at sub-6 GHz use signals broadcasted (quasi) omnidirectionally, which are not subject to deafness. A synchronization signal sent by \gls{lte} base stations, for example, can be received by all the users in their coverage area, simultaneously. In \gls{5g} networks with \gls{mmwave} communications, however, this paradigm has changed, as directionality prevents devices and base stations from transmitting and receiving omnidirectionally. Such systems, indeed, need beamforming to improve the link budget and extend the possible communication range~\cite{giordani2018tutorial}. To achieve the maximum gain, the endpoints need to align their transmit and receive beams, and this may introduce delays in the link setup, limit the awareness of the spectrum utilization, and impair the reliability of communications in highly mobile scenarios.

\subsection{Link Budget and Initial Access Latency Comparison}

Fig.~\ref{fig:snr} compares the \gls{snr} at different distances (5, 30 and 100 m) for mmWave and terahertz links. The \gls{snr} is given by the ratio between the received signal power and the noise power, without the beamforming gain, to analyze the impact of the higher carrier frequency on the propagation loss. The latter is computed using established models, i.e., the 3GPP model in an urban canyon scenario in the frequency range considered for 3GPP NR~\cite{38901}, and~\cite{jornet2011thz} for terahertz links in the 300-1000 GHz spectrum. The bandwidth is 400 MHz for \glspl{mmwave}, i.e., the maximum bandwidth per carrier of 3GPP NR, and 50 GHz for terahertz, compliant with IEEE 802.15.3d.

For example, we observe that the \gls{snr} gap between a carrier at 30 GHz and one at 430 GHz is 37 dB. 
As mentioned in Sec.~\ref{sec:intro}, this difference can be compensated for by increasing the number of antenna elements in each node, thanks to the smaller wavelength at terahertz when compared to \glspl{mmwave} and sub-6 GHz networks. This translates into narrower beams, which improve the link budget but, at the same time, increase the deafness problem, limiting the awareness of mobile nodes.

The impact of highly directional antennas on the \gls{mac} and higher layers is shown in Fig.~\ref{fig:ia}, which reports the latency for the initial link establishment, with an exhaustive scan to identify the best transmit and receive beam, using the frame structure of 3GPP NR and the latency model from~\cite{giordani2018tutorial}. Notably, the base stations send 8, 16, 32 or 64 directional synchronization signals every $T_{ss}$ seconds, which allow mobile nodes to assess the channel quality and decide on the best beam pair. For the mmWave links, we consider a setup with 16 antennas at the base stations and 4 at the \glspl{ue}. This yields a gain of up to 18 dB, which is enough to have a \gls{los} \gls{snr} in the order of 0 dB at 100 m and 52.6 GHz. For terahertz, conversely, the configuration of Fig.~\ref{fig:ia} is designed to match the overall link budget of the mmWave setup, with 1024 elements at the base stations, and 256 at the mobile endpoint, and a gain of up to 54 dB. While the link budget is the same, the tighter beam requires more directional synchronization signals for the same coverage area. Therefore, \emph{an exhaustive search at the transmit and receive side on the 3GPP NR frame structure is slow but still feasible at \glspl{mmwave}} (with a delay smaller than 1 s in the \textit{worst} configuration), but \emph{is not a viable option for terahertz, which would require more than 32 s to complete the scan in the \textit{best} case.}

\subsection{Beam Operations for Terahertz}

Consequently, beam management protocols for the terahertz \gls{mac} layer have to carefully consider the trade-off between improving the link budget and the need for awareness of spectrum and infrastructure. Research on beam management solutions for \glspl{mmwave} has already proposed several alternatives to a basic exhaustive search, which could be partially adapted to the terahertz domain~\cite{akyildiz2018combating}. Additionally, the characteristics of the terahertz spectrum could be exploited to improve directional operations~\cite{xia2019link}. Promising research directions will involve a combination of:
\begin{itemize}
	\item a redesign of the frame structure with respect to that of NR considered in Fig.~\ref{fig:ia}, and of synchronization and reference signals, to exploit the larger bandwidth available at terahertz. Shorter pilots could result in more opportunities to transmit synchronization signals without impacting the control overhead, and multiple user-specific tracking signals could be multiplexed in frequency, if the link budget for the interested devices is high enough;
	\item advanced antenna architectures could be exploited to transmit and receive directional signals from multiple transceivers at the same time (e.g., with digital beamforming), or to infer angle of departure and arrival of terahertz signals (e.g., with leaky-wave antennas~\cite{ghasempour2020single}). Further research efforts are required to design and realize digital architectures and antennas for the terahertz spectrum, but promising results have also been obtained with multi-beam solutions based on plasmonic nano-antennas~\cite{AKYILDIZ201646}, or by considering the information gathered not only through the main beam, but also with sidelobes;
	\item multi-stage beam management schemes, where beam configurations with different beam widths and, consequently, gains are used for different steps of either tracking or channel sensing procedures;
	\item context-based schemes, which use additional information to gather awareness of the surrounding environment. Notably, at terahertz, the large bandwidth and the propagation characteristics make the medium particularly suitable for radio-frequency sensing~\cite{rappaport2019wireless}. Therefore, the same radio interface could be used to transmit data and to map the surrounding environment, identifying, for example, sources of blockage;
	\item \gls{mac} protocols that rely on multi-connectivity, i.e., the availability of multiple radios in the mobile devices and base stations, to exploit different frequency bands for different tasks: sub-6 GHz links can provide a control overlay, and assist beam management operations at terahertz with reliable feedback links.
\end{itemize}

\section{Link Design for Ultra-High Bandwidth}
\label{sec:link}

The unprecedented availability of bandwidth at terahertz, as shown in Fig.~\ref{fig:challenges}, and the highly directional transmissions also call for further research on medium access and retransmission mechanisms. Notably, the high physical layer rate may allow MAC protocols to still achieve high throughput, while
accepting a lower bandwidth/spectral efficiency in favor of simplicity. Indeed, protocols which require a high level of coordination may be impractical for terahertz links.

For the medium access, the main design choice is between scheduled and contention-based. Each of these paradigms presents potentials and limitations which could be exploited for different use cases and scenarios. In particular,
\begin{itemize}
	\item with scheduled access, the base stations of a cellular network allocate time and frequency resources to the connected users, avoiding collisions and, consequently, increasing spectrum efficiency. Additionally, the centralized control at the base stations enables prompt link adaptation and tracking. Indeed, the base stations can allocate specific tracking or reference signals to connected users, and coordinate with the infrastructure to understand the evolution of the channel dynamics for a certain user. However, it may be challenging to establish and maintain connectivity towards the fixed infrastructure, given the harsh propagation environment in the terahertz band, and the control plane needs to cope with directionality;
	\item most research related to terahertz \gls{mac} layer design has proposed approaches based on contention-based medium access~\cite{hossain2018terasim}. With respect to a scheduled \gls{mac}, a contention-based \gls{mac} does not need control plane connectivity toward the infrastructure. Moreover, the highly directional transmissions and their short duration (thanks to the ultra-high available data rate) limit the impact of collisions among different concurrent data exchanges. Nonetheless, the two endpoints of the communication still need to agree upon the optimal beam pair, thus a beam search step needs to be included at every channel access~\cite{xia2019link}.
\end{itemize}

For retransmissions, unpredictable blockers and frequent beam or base station updates could disrupt the constant flow of acknowledgments between the receiver and the transmitter, thus impacting the design of retransmission policies. Additionally, given the high datarate, coding for forward error correction could make the transceiver design overly complex. Therefore, an efficient retransmission process remains an open issue. To this end, network coding techniques~\cite{phung2020improcing} have low complexity implementations and simplify the retransmission process, as they do not require the retransmission of a specific packet, but rather of a random combination of the batch of packets that the receiver needs to decode.

\section{Making the Network Scale}
\label{sec:scale}

\gls{mmwave} base stations are being deployed as high-density small cells in 5G networks, since their coverage is limited~\cite{akyildiz2018combating}. Following this trend, and considering the higher pathloss in the terahertz spectrum, the density of a cellular network operating at terahertz will likely  increase even further. This will have an impact on the capital and operating expenditures, the energy consumption, and the complexity and scale of the backhaul infrastructure. In this section, we will review the main challenges and research directions associated to the deployment and operation of a terahertz network at a massive scale.

\begin{figure}[t]
	\centering
	\setlength\fwidth{.8\columnwidth}
	\setlength\fheight{0.5\columnwidth}
	\input{figures/coverage.tex}
	\caption{Probability of having an SNR higher than 0 dB for different deployment configurations. The mmWave channel is modeled according to~\cite{38901}, i.e., through a combination of probabilistic \gls{los}  and path loss models for an urban scenario with microcells. For terahertz, the path loss of a \gls{los}-only channel is modeled with~\cite{jornet2011thz}. The other link parameters are as in Fig.~\ref{fig:snr}.}
	\label{fig:coverage}
\end{figure}
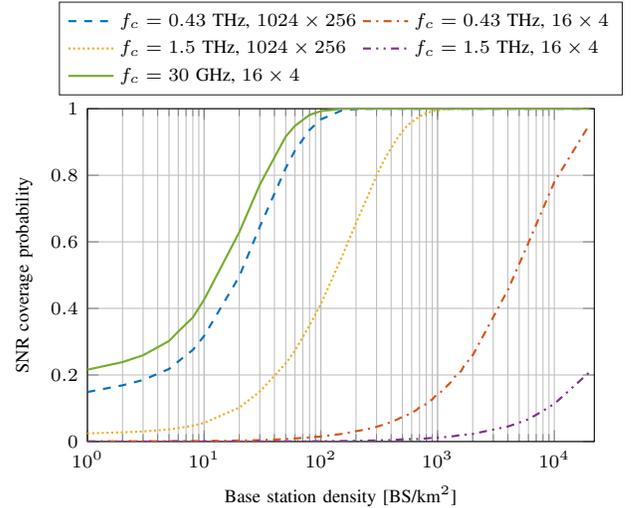

Figure~\ref{fig:coverage} illustrates the differences in deployment density that can be expected for \gls{mmwave} and terahertz systems. We consider Monte Carlo simulations, in which macro base stations are randomly deployed outdoors according to a Poisson point process, and with the coverage probability computed as the probability of having an \gls{snr} above a threshold of 0 dB between a test user, at the center of the deployment, and at least one base station. For example, this can be a metric that is measured by the mobile devices during an initial access attempt~\cite{giordani2018tutorial}. The path loss and beamforming gain are modeled as in Fig.~\ref{fig:link}. We notice that with 16 antenna elements at the base station and 4 at the mobile device, the 30 GHz network has a coverage probability higher than 0.95 with 60 base stations per km$^2$, which corresponds to an average cell radius of 72 m. The same antenna configuration does not guarantee an adequate performance at terahertz, since a similar coverage probability requires $2\cdot10^4$ BS/km$^2$ at 0.43 THz. As discussed in Sec.~\ref{sec:awareness}, antenna arrays with a larger number of antenna elements will be necessary when considering these frequency bands. With 1024 antenna elements for the base stations and 256 for the mobile devices, it is possible to reach a coverage probability of 0.95 with at least 100 BS/km$^2$ at 0.43 THz, and 600 BS/km$^2$ at 1.5 THz, with an average cell radius of 56 and 23 m, respectively. 

\subsection{Energy-aware Network Design}

6G will further improve the energy efficiency of 5G networks~\cite{giordani20196g}, to balance the higher number of terahertz nodes to be powered up than at \glspl{mmwave} or sub-6 GHz. Possible strategies toward the reduction of energy consumption at terahertz include:
\begin{itemize}
	\item a lean control plane design, that minimizes the control messages and always-on signals for control operations;
	\item quick sleep cycles for base stations in dense clusters with low traffic. With the deployment density foreseen in Fig.~\ref{fig:coverage}, the number of base stations will likely approach, and, in some areas, may even exceed that of active users in the network. Therefore, terahertz base stations should  quickly turn off the main radio functionalities (to save energy) when there are no connected users. This will require coordination among neighboring base stations, which can notify each other of users relocating to switched-off base stations, so that they can swiftly resume operations, or wake-up radio systems, exploiting multi connectivity with low-power radios;
	\item energy-saving states for mobile devices without active transmissions, alternating stand-by with intervals in which the network could page the device. A specific challenge for terahertz is maintaining connectivity in stand-by with directional links, as the best beam pair could change if the mobile device in a low-energy mode moves through the network;
	\item energy harvesting, with the infrastructure and the mobile devices exploiting modern harvesting circuitry to self-sustain during sleep cycles.
\end{itemize}

\subsection{A Control Plane for Reflecting Arrays and Metasurfaces}

Even with large antenna arrays, uniform access coverage with terahertz links may be infeasible, in terms of cost and energy consumption. Hotspots and indoor scenarios are more likely candidates for early terahertz deployments. For these, the traditional base stations coverage can be enhanced with new network infrastructure elements, namely, reflecting arrays and metasurfaces~\cite{akyildiz2018combating}. These devices are based, respectively, on phased arrays and nanomaterials that can steer the terahertz wave impinging on them, thus reflecting the signal transmitted by a terahertz node towards a mobile user. Their deployment improves the link budget and the coverage in \gls{nlos} conditions, and reduces the density of full base stations. However, their integration with the fixed infrastructure requires the design of a dedicated control plane, with protocols and networking procedures to manage, among others, the handoff of users across different reflecting devices, and the tracking in highly mobile scenarios. Moreover, the scale of terahertz networks will make manual configuration impractical, calling for intelligent procedures that automatically connect and jointly optimize the parameters of base stations and reflecting nodes in a plug-and-play fashion.

\section{Managing the Spectrum}
\label{sec:coordination}

The large, untapped portions of spectrum at terahertz also create new opportunities for spectrum management. Indeed, besides allocating large bandwidths to the radio access, it is possible to enhance the network by deploying novel spectrum reuse schemes, in-band, high-capacity wireless backhaul and multi connectivity. 

\subsection{Interference}

First of all, it is important to characterize the impact of interference when considering the extremely directional links~\cite{petrov2017interference}. Indeed, as for \glspl{mmwave}, terahertz networks can be noise-limited thanks to beamformed transmissions and the limited coverage of each base station. However, at the same time, the high density that is needed to provide coverage (as discussed in Sec.~\ref{sec:scale}) may introduce additional interference. Finally, the bursty transmissions complicate the tracking and prediction of possible interference sources. To this end,~\cite{barazideh2020reinforcement} proposes a reinforcement learning scheme to detect and mitigate intermittent interference sources, showing how adaptive, learning-based approaches can cope with challenging and dynamic terahertz scenarios.
The behavior of interference at terahertz can affect the design of spectrum reuse and sharing schemes, which may be tuned as more or less aggressive according to the need to isolate from cross-cell interference. Additionally, interference management strategies should be re-designed to account for both active and passive users (e.g., to protect incumbents that use the spectrum for radio astronomy and Earth atmospheric science), with coordination loops across terahertz nodes which need to be fast enough to address a highly dynamic interference environment.

\subsection{Wireless Backhaul}

The large available bandwidth can also be used for in-band backhaul for terahertz base stations. The high deployment density will make wired backhaul to each base station extremely expensive, thus calling for a fully wireless solution, e.g., as proposed with \gls{iab} in 5G networks. Compared to \gls{iab} at \glspl{mmwave}, terahertz benefits from the higher spectrum availability, which could improve the quality of service for end-to-end traffic flows over multiple wireless hops. Moreover, fixed relays simplify beam management, but network operators should carefully design the deployment and the topology of the wireless backhaul network to provide connectivity to all the base stations in the presence of blockage.

\subsection{Multi connectivity}

Finally, 6G networks will rely on a combination of sub-6 GHz, \gls{mmwave} and terahertz bands, and, possibly, optical wireless links~\cite{giordani20196g,rappaport2019wireless}. The network infrastructure and the mobile devices will therefore need to nimbly adapt and use the carrier that provides the best performance. 

\begin{figure}[t]
	\centering
	\setlength\fwidth{.8\columnwidth}
	\setlength\fheight{0.45\columnwidth}
	\input{figures/udp-throughput-thz.tex}
	\caption{UDP throughput with different source rates $R$ and distances between the base stations and mobile devices.}
	\label{fig:udp}
\end{figure}
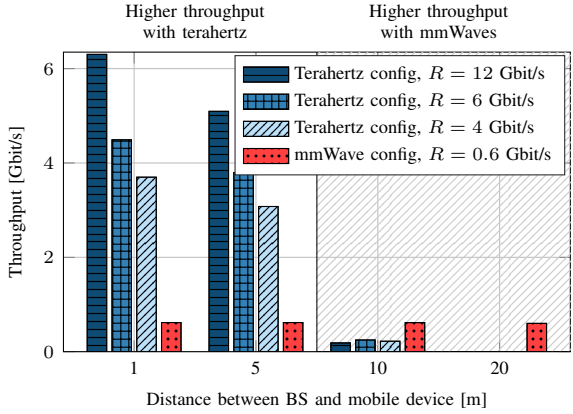

In Fig.~\ref{fig:udp} we compare the throughput of a terahertz link at $1.0345$ THz, with $74$ GHz of bandwidth (corresponding to the first available window in the spectrum above $1$ THz~\cite{hossain2018terasim}), and a mmWave link operating at $28$ GHz with the maximum bandwidth allowed in 3GPP NR (i.e., $400$ MHz), through full-stack simulations in ns-3 with the TeraSim~\cite{hossain2018terasim} and mmWave modules~\cite{mezzavilla2018end}. The application is a constant-bitrate source, with UDP at the transport layer. The wireless nodes are equipped with directional arrays, i.e., a rotating array for terahertz, as in the default \textit{macro} scenario of TeraSim~\cite{hossain2018terasim}, and phased arrays with 64 and 16 antenna elements at the base station and the device~\cite{mezzavilla2018end}.

This network configuration highlights two operating regimes, according to the distance between the base station and the mobile device. For short distances, i.e., 1 and 5 m, the terahertz link benefits from the larger bandwidth with respect to the mmWave connection, and provides a higher throughput. However, as the distance increases, the pathloss for the terahertz link increases at a faster rate than for the mmWave one, which eventually becomes the better choice in terms of throughput at 10 and 20 m. Therefore, 6G terahertz devices should exploit multi connectivity not only for the control plane and beam management, as mentioned in Sec.~\ref{sec:awareness}, but also for the user plane, forwarding data packets on the different available radio interfaces to provide diversity.

\section{Moving bits end to end}
\label{sec:e2e}

As discussed in Sec.~\ref{sec:intro}, terahertz links in cellular or ad hoc networks will constitute only a fraction of the hops in an end-to-end connection, and will carry traffic generated by a wide range of different applications, with various underlying transport protocols (e.g., TCP). The end-to-end performance is thus determined by the interaction between these applications and protocols and the resources at the physical layer, and a sub-optimal interplay may prevent full exploitation of the large bandwidth and data rates of terahertz connections. Prior research has shown that, at \glspl{mmwave}, the highly intermittent channel and beamforming degrade the performance of traditional TCP congestion control schemes~\cite{mezzavilla2018end}. Therefore, the interplay with the transport layer should be considered when designing the protocol stack for terahertz links as well.

\begin{figure}[t]
	\centering
	\setlength\fwidth{.8\columnwidth}
	\setlength\fheight{0.45\columnwidth}
	\input{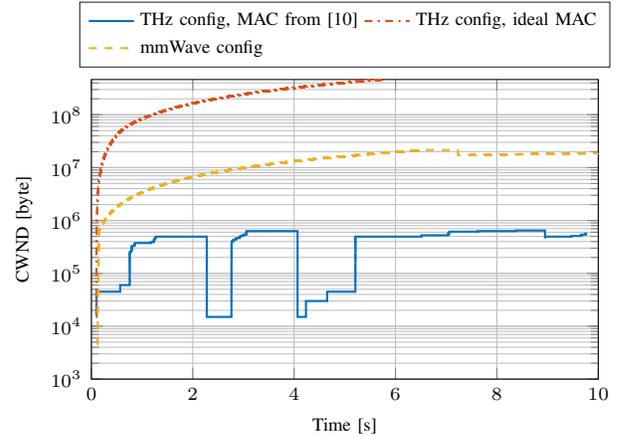}
	\caption{Evolution of the TCP CUBIC congestion window. The base station and mobile device are at a distance of 1 m, with the same configuration as in Fig.~\ref{fig:udp}.}
	\label{fig:tcp}
\end{figure}

Figure~\ref{fig:tcp} exemplifies the pitfalls of TCP for terahertz links by comparing the evolution of the congestion window for a single TCP flow on a mmWave link (28 GHz), with a scheduled \gls{mac}~\cite{mezzavilla2018end,giordani2018tutorial}, and a terahertz link (1.0345 THz) with two \gls{mac} layer configurations. The first, from~\cite{xia2019link}, has beam management and contention, while the other is ideal, i.e., all the resources are always allocated to the same user with the best beam pair. We observe that the congestion window with the terahertz link and the realistic \gls{mac} is reduced multiple times, not because capacity is reached, but because of the inefficient interplay between the contention-based access and the TCP timers, that triggers timeouts and congestion recovery. Overall, this configuration performs much worse than the mmWave one, despite the larger bandwidth, with an average throughput of 66 Mbit/s vs 520 Mbit/s. The ideal \gls{mac} configuration, instead, highlights another issue that TCP may suffer from on terahertz links, i.e., the sub-optimal use of the available physical layer resources due to the slow linear ramp-up of TCP in congestion avoidance.

Finally, current protocol stacks in mobile devices are not designed to handle data rates that can reach tens of gigabits per second. As the throughput increases, the CPU of the device becomes busier processing the received packets. Moreover, congestion and flow control decisions have to be made much more frequently. Therefore, the networking stack processing may quickly deplete the battery of mobile devices when operating at high throughput. This makes the case for additional simulation-based and experimental research on the design and performance of \textit{simpler} network and transport protocols at terahertz,  e.g., based on UDP, QUIC, or on future evolutions of TCP. Moreover, further analysis is needed to understand whether congestion control mechanisms are useful for terahertz links, considering the datarates at stake.

\section{Conclusions and Future Work}
\label{sec:concl}

The terahertz physical layer design still presents many open research challenges. This paper focused on key issues for the higher layers of the protocol stack, also discussing deployment and energy-related challenges.
We identified relevant networking problems, providing insights and preliminary results that can drive future research on 6G terahertz networks at the \gls{mac} layer and above. Five key research questions emerged from our analysis:
\begin{enumerate}
	\item How can \textit{beam management} and medium access schemes be designed to exploit the characteristics of the \textit{terahertz spectrum}, e.g., by combining sensing, large antenna arrays, and communications?
	\item Do terahertz networks need a \textit{reliable control plane} and a scheduled MAC, or does contention-based access provide a better trade off between \textit{complexity and performance}? How can the control plane be extended to metasurfaces?
	\item Which are the most effective strategies to deploy a \textit{high-density, energy-efficient} network? 
	\item Which policies can be developed to manage the terahertz spectrum, considering dynamic \textit{interference} sources, the large available bandwidth, and the possibility of using \textit{multi connectivity}? 
	\item How can \textit{transport protocol designs and implementations evolve} to satisfy the requirements of ultra-high bandwidth, highly variable links?
\end{enumerate}

These research questions, and the insights we provided in the paper, could be used as a starting point to further progress the full-stack, end-to-end analysis and design of terahertz networks, to fully profit from the unprecedented amount of bandwidth available in this portion of the spectrum.

\section*{Acknowledgements}
The work of M. Polese and T. Melodia was supported in part by the US Office of Naval Research under Grant N00014-20-1-2132. The work of J. Jornet was supported in part by the US National Science Foundation Grant CNS-2011411.

\bibliographystyle{IEEEtran}
\bibliography{bibl.bib}

\begin{IEEEbiographynophoto}{Michele Polese}
[M'20] is a research scientist at Northeastern University, Boston. He obtained his Ph.D. from the University of Padova, Italy, in 2020, where he also was a postdoctoral researcher and adjunct professor. He visited NYU, AT\&T Labs, and Northeastern University. His research focuses on protocols and architectures for future wireless networks.
\end{IEEEbiographynophoto}%

\begin{IEEEbiographynophoto}{Josep Miquel Jornet} [M'13] received the Ph.D. degree in Electrical and Computer Engineering (ECE) from the Georgia Institute of Technology in 2013. Between 2013 and 2019, he was with the Department of Electrical Engineering at the University at Buffalo. Since August 2019, he has been an Associate Professor in the Department of ECE at Northeastern University. His research interests are in Terahertz-band communications and Wireless Nano-bio-communication Networks. He has co-authored more than 120 peer-reviewed scientific publications, one book, and has been granted 3 US patents. He is serving as the lead PI on multiple grants from U.S. federal agencies.
\end{IEEEbiographynophoto}%

\begin{IEEEbiographynophoto}{Tommaso Melodia}
[F’18] received the Ph.D. degree in Electrical and Computer Engineering from the Georgia Institute of Technology in 2007. He is the William Lincoln Smith Professor at Northeastern University, the Director of the Institute for the Wireless Internet of Things, and the Director of Research for the PAWR Project Office. His research focuses on modeling, optimization, and experimental evaluation of wireless networked systems. He serves as Editor in Chief for Computer Networks.
\end{IEEEbiographynophoto}%

\begin{IEEEbiographynophoto}{Michele Zorzi}
[F'07] is with the Information Engineering Department of the University of Padova, focusing on wireless communications research. He was Editor-in-Chief of IEEE Wireless Communications from 2003 to 2005, IEEE Transactions on Communications from 2008 to 2011, and IEEE Transactions on Cognitive Communications and Networking from 2014 to 2018. He served ComSoc as a Member-at-Large of the Board of Governors from 2009 to 2011, as Director of Education and Training from 2014 to 2015, and as Director of Journals from 2020 to 2021.
\end{IEEEbiographynophoto}

\end{document}

%% file: figures/initial_access_nr_thz.tex
%
%
\definecolor{mycolor1}{rgb}{0.00000,0.44700,0.74100}%
\definecolor{mycolor2}{rgb}{0.85000,0.32500,0.09800}%
\definecolor{mycolor3}{rgb}{0.92900,0.69400,0.12500}%
\definecolor{mycolor4}{rgb}{0.49400,0.18400,0.55600}%
\begin{tikzpicture}
\pgfplotsset{
tick label style={font=\scriptsize},
label style={font=\scriptsize},
legend  style={font=\scriptsize}
}

\begin{axis}[%
width=0.951\fwidth,
height=\fheight,
at={(0\fwidth,0\fheight)},
scale only axis,
xmin=7,
xmax=65,
xlabel={Number of synch signals in $T_{ss}$},
ymode=log,
ymin=0.01,
ymax=1200,
ylabel shift=-5pt,
yminorticks=true,
ylabel={Full-scan Initial Access Delay [s]},
axis background/.style={fill=white},
xmajorgrids,
ymajorgrids,
yminorgrids,
legend style={legend cell align=left, align=left, draw=white!15!black},
xtick=data,
]
\addplot [color=mycolor1, line width=1.2pt, mark=o, mark options={solid, mycolor1}, dashdotted]
  table[row sep=crcr]{%
8	0.52005358125\\
16	0.26005358125\\
32	0.12055358125\\
64	0.06055358125\\
};

\addplot [color=mycolor1, line width=1.2pt, mark=asterisk, mark options={solid, mycolor1}]
  table[row sep=crcr]{%
8	1028.84011608125\\
16	514.42011608125\\
32	257.20061608125\\
64	128.60061608125\\
};

\addplot [color=mycolor2, line width=1.2pt, mark=o, mark options={solid, mycolor2}, dashdotted]
  table[row sep=crcr]{%
8	0.13005358125\\
16	0.06505358125\\
32	0.03055358125\\
64	0.01555358125\\
};

\addplot [color=mycolor2, line width=1.2pt, mark=asterisk, mark options={solid, mycolor2}]
  table[row sep=crcr]{%
8	257.21011608125\\
16	128.60511608125\\
32	64.30061608125\\
64	32.15061608125\\
};

\node (centerThz) at (32, 120) {};

\node (centerMm) at (50, 0.05) {};

\end{axis}

\begin{axis}[%
width=0.951\fwidth,
height=\fheight,
at={(0\fwidth,0\fheight)},
scale only axis,
xmin=7,
xmax=65,
ymode=log,
ymin=0.01,
ymax=600,
hide x axis,
hide y axis,
legend style={legend cell align=left, align=left, draw=white!15!black, at={(.98,1.02)}, anchor=north east}
]
\addplot [color=mycolor1, line width=1.2pt]
  table[row sep=crcr]{%
-2	2\\
-3  4\\
};
\addlegendentry{$T_{ss} =20$ ms}

\addplot [color=mycolor2, line width=1.2pt]
  table[row sep=crcr]{%
-2	2\\
-3  4\\
};
\addlegendentry{$T_{ss} =5$ ms}

\end{axis}




\draw[black, thick] (centerThz) ellipse (0.2cm and 0.6cm) node[anchor=north, align=center, font=\tiny, yshift=-.7cm] {\scriptsize Terahertz configuration ($\sim 54$ dB gain)\\1024 elements at the base station, 256 at the mobile node};

\draw[black, thick, dashdotted] (centerMm) ellipse (0.2cm and 0.6cm) node[anchor=south, align=center, font=\tiny, yshift=.7cm, xshift=-.65cm] {\scriptsize mmWave configuration ($\sim 18$ dB gain)\\16 elements at the base station, 4 at the mobile node};

\end{tikzpicture}%

%% file: figures/coverage.tex
%
%
\definecolor{mycolor1}{rgb}{0.00000,0.44700,0.74100}%
\definecolor{mycolor2}{rgb}{0.85000,0.32500,0.09800}%
\definecolor{mycolor3}{rgb}{0.92900,0.69400,0.12500}%
\definecolor{mycolor4}{rgb}{0.49400,0.18400,0.55600}%
\definecolor{mycolor5}{rgb}{0.46600,0.67400,0.18800}%
\begin{tikzpicture}
\pgfplotsset{every tick label/.append style={font=\scriptsize}}

\begin{axis}[%
xmode=log,
width=0.951\fwidth,
height=\fheight,
at={(0\fwidth,0\fheight)},
scale only axis,
xmin=1,
xmax=22000,
ymin=0,
ymax=1,
axis background/.style={fill=white},
xmajorgrids,
ymajorgrids,
ylabel={SNR coverage probability},
ylabel style={font=\scriptsize},
xlabel={Base station density [BS/km$^2$]},
xlabel style={font=\scriptsize},
legend columns=2,
xminorgrids,
legend style={legend cell align=left, align=left, draw=white!15!black, font=\scriptsize, at={(0.5, 1.04)}, anchor=south}
]
\addplot [color=mycolor1, thick, dashed]
  table[row sep=crcr]{%
1	0.14882\\
2	0.16896\\
3	0.18484\\
5	0.21798\\
6	0.2406\\
8	0.2752\\
10	0.31728\\
20	0.49594\\
30	0.64536\\
50	0.82158\\
60	0.87414\\
80	0.93612\\
100	0.96774\\
150	0.9943\\
200	0.9991\\
250	0.99982\\
300	0.99998\\
350	0.99998\\
400	1\\
450	1\\
500	1\\
550	1\\
600	1\\
650	1\\
700	1\\
750	1\\
800	1\\
850	1\\
900	1\\
950	1\\
1000	1\\
1100	1\\
1200	1\\
1300	1\\
1400	1\\
1500	1\\
1600	1\\
1700	1\\
1800	1\\
1900	1\\
2000	1\\
4000	1\\
6000	1\\
8000	1\\
10000	1\\
20000	1\\
};
\addlegendentry{$f_c = 0.43$ THz, $1024\times256$}

\addplot [color=mycolor2, thick, dashdotted]
  table[row sep=crcr]{%
1	0.00058\\
2	0.00074\\
3	0.00086\\
5	0.00102\\
6	0.00156\\
8	0.00154\\
10	0.00156\\
20	0.00314\\
30	0.00418\\
50	0.00738\\
60	0.00924\\
80	0.01232\\
100	0.01532\\
150	0.02276\\
200	0.03098\\
250	0.03722\\
300	0.0443\\
350	0.05158\\
400	0.05862\\
450	0.0663\\
500	0.0743\\
550	0.07974\\
600	0.08638\\
650	0.0932\\
700	0.10236\\
750	0.10942\\
800	0.11254\\
850	0.12176\\
900	0.12596\\
950	0.13608\\
1000	0.13968\\
1100	0.15328\\
1200	0.1663\\
1300	0.17844\\
1400	0.194\\
1500	0.20314\\
1600	0.21272\\
1700	0.23064\\
1800	0.23892\\
1900	0.2528\\
2000	0.25922\\
4000	0.45548\\
6000	0.59688\\
8000	0.69982\\
10000	0.77972\\
20000	0.9521\\
};
\addlegendentry{$f_c = 0.43$ THz, $16\times4$}

\addplot [color=mycolor3, thick, densely dotted]
  table[row sep=crcr]{%
1	0.0245\\
2	0.02756\\
3	0.03038\\
5	0.03634\\
6	0.04034\\
8	0.04754\\
10	0.05604\\
20	0.10192\\
30	0.14992\\
50	0.2346\\
60	0.27322\\
80	0.34884\\
100	0.41296\\
150	0.55146\\
200	0.65626\\
250	0.73386\\
300	0.79944\\
350	0.84712\\
400	0.88152\\
450	0.91032\\
500	0.9321\\
550	0.94922\\
600	0.96018\\
650	0.96838\\
700	0.97656\\
750	0.9821\\
800	0.98656\\
850	0.98902\\
900	0.99236\\
950	0.99402\\
1000	0.99496\\
1100	0.99766\\
1200	0.99824\\
1300	0.99922\\
1400	0.99948\\
1500	0.99948\\
1600	0.9999\\
1700	0.99982\\
1800	1\\
1900	1\\
2000	1\\
4000	1\\
6000	1\\
8000	1\\
10000	1\\
20000	1\\
};
\addlegendentry{$f_c = 1.5$ THz, $1024\times256$}

\addplot [color=mycolor4, thick, dashdotdotted]
  table[row sep=crcr]{%
1	4e-05\\
2	4e-05\\
3	8e-05\\
5	8e-05\\
6	8e-05\\
8	8e-05\\
10	8e-05\\
20	0.00014\\
30	0.00026\\
50	0.00046\\
60	0.00092\\
80	0.00062\\
100	0.00108\\
150	0.00162\\
200	0.0024\\
250	0.00292\\
300	0.0033\\
350	0.00412\\
400	0.00438\\
450	0.00534\\
500	0.0068\\
550	0.00652\\
600	0.00772\\
650	0.00702\\
700	0.00786\\
750	0.00834\\
800	0.00938\\
850	0.01016\\
900	0.01094\\
950	0.01158\\
1000	0.01144\\
1100	0.0126\\
1200	0.014\\
1300	0.01512\\
1400	0.01662\\
1500	0.01832\\
1600	0.0179\\
1700	0.01974\\
1800	0.02196\\
1900	0.02228\\
2000	0.02292\\
4000	0.04506\\
6000	0.06674\\
8000	0.0923\\
10000	0.11318\\
20000	0.2084\\
};
\addlegendentry{$f_c = 1.5$ THz, $16\times4$}

\addplot [color=mycolor5, thick]
  table[row sep=crcr]{%
1	0.21584\\
2	0.2387\\
3	0.25928\\
5	0.30196\\
6	0.33052\\
8	0.37274\\
10	0.42626\\
20	0.6289\\
30	0.77188\\
50	0.9163\\
60	0.9485\\
80	0.98132\\
100	0.99282\\
150	0.99934\\
200	0.99998\\
250	1\\
300	1\\
350	1\\
400	1\\
450	1\\
500	1\\
550	1\\
600	1\\
650	1\\
700	1\\
750	1\\
800	1\\
850	1\\
900	1\\
950	1\\
1000	1\\
1100	1\\
1200	1\\
1300	1\\
1400	1\\
1500	1\\
1600	1\\
1700	1\\
1800	1\\
1900	1\\
2000	1\\
4000	1\\
6000	1\\
8000	1\\
10000	1\\
20000	1\\
};
\addlegendentry{$f_c = 30$ GHz, $16\times4$}

\end{axis}
\end{tikzpicture}%

%% file: figures/udp-throughput-thz.tex
\begin{tikzpicture}
\pgfplotsset{every tick label/.append style={font=\scriptsize}}

\definecolor{mycolor1}{rgb}{0.00000,0.44700,0.74100}%
\definecolor{mycolor2}{rgb}{0.6784,0.8471,0.9020}%
\definecolor{mycolor4}{rgb}{0,0.44710,0.74120}%
\definecolor{mycolor5}{rgb}{1.0000,0.6275,0.4784}%
\definecolor{mycolor3}{rgb}{0.92900,0.69400,0.12500}%
\definecolor{mycolor6}{rgb}{1.0000,0.9804,0.8039}%
\definecolor{mycolor21}{rgb}{0.83529411764,0.92156862745,0.94509803921}
\definecolor{mycolor11}{rgb}{0.50196078431,0.72156862745,0.86666666666}
\definecolor{mycolor31}{rgb}{0.96470588235,0.83921568627,0.5725490196}

\definecolor{blue1}{rgb}{0.058,0.298,0.459}
\definecolor{blue2}{rgb}{0.196,0.509,0.721}
\definecolor{blue3}{rgb}{0.733,0.882,0.98}
\definecolor{coralred}{rgb}{1,0.251,0.251}

\begin{axis}[
width=0.951\fwidth,
height=\fheight,
at={(0\fwidth,0\fheight)},
scale only axis,
bar shift auto,
xmin=0.5, 
xmax=4.5,
ymin=0, 
ymax=6.350,
axis background/.style={fill=white},
xmajorgrids,
ymajorgrids,
xminorgrids,
yminorgrids,
ylabel={Throughput [Gbit/s]},
xlabel={Distance between BS and mobile device [m]},
xlabel style={font=\scriptsize},
ylabel style={font=\scriptsize},
legend style={legend cell align=left, align=left, draw=white!15!black, font=\scriptsize, at={(0.99, 0.99)}, anchor=north east},
xtick=data,
xticklabels={1, 5, 10, 20},
enlarge x limits=0.02,
]
















\filldraw [pattern=north east lines,pattern color=black!20] (2.5,0) rectangle (5,7);

\addplot [ybar, bar width=0.16, fill=blue1, draw=black, area legend, postaction={pattern=horizontal lines}]
table {%
1 6.30237417101068
2 5.09487020564664
3 0.185685666441762
4 0.000183952172684211
};
\addlegendentry{Terahertz config, $R=12$ Gbit/s}

\addplot [ybar, bar width=0.16, fill=blue2, draw=black, area legend, postaction={pattern=grid}]
table {%
1 4.49130371349281
2 3.79839313290088
3 0.247776221336205
4 0.0000960620510756655
};
\addlegendentry{Terahertz config, $R=6$ Gbit/s}

\addplot [ybar, bar width=0.16, fill=blue3, draw=black, area legend, postaction={pattern=north east lines}]
table {%
1 3.69893835580493
2 3.0778277553859
3 0.221248728294914
4 0.0000441566775723493
};
\addlegendentry{Terahertz config, $R=4$ Gbit/s}

\addplot [ybar, bar width=0.16, fill=coralred, draw=black, area legend, postaction={pattern=dots}]
table {%
1 0.618067302712264
2 0.618067302712264
3 0.617067302712264
4 0.600067302712264
};
\addlegendentry{mmWave config, $R=0.6$ Gbit/s}

\node (mmw) at (3.5,7) {};
\node (thz) at (1.5,7) {};

\end{axis}

\node[align=center, font=\scriptsize] at (mmw) {Higher throughput\\with \glspl{mmwave}};

\node[align=center, font=\scriptsize] at (thz) {Higher throughput\\with terahertz};

\end{tikzpicture}